\documentclass[12pt]{iopart}

\usepackage[english]{babel}
\usepackage[utf8x]{inputenc}
\usepackage[T1]{fontenc}
\usepackage{amsfonts}
\usepackage{amssymb}
\usepackage{epigraph}
\usepackage{float}

\usepackage{bm}

\usepackage{xcolor}

\usepackage{tikz}
\usetikzlibrary{calc}

\usepackage[a4paper,top=3cm,bottom=2cm,left=3cm,right=3cm,marginparwidth=1.75cm]{geometry}

\usepackage{graphicx}
\usepackage[colorlinks=true, allcolors=blue]{hyperref}

\newcommand{\la}{\label}
\newcommand{\bbm}{\begin{multline}}
\newcommand{\eem}{\end{multline}}
\newcommand{\be}{\begin{equation}}
\newcommand{\ee}{\end{equation}}
\newcommand{\bea}{\begin{eqnarray}}
\newcommand{\eea}{\end{eqnarray}}

\begin{document}


\title[The boundary density profile of a Coulomb droplet]{The boundary density profile of a Coulomb droplet.\\ Freezing at the edge. }


\author{Gabriel Cardoso}
\address{Department of Physics and Astronomy, Stony Brook University, Stony Brook, NY 11794, USA}

\author{Jean-Marie St\'ephan}
\address{Univ Lyon, CNRS, Universit\'e Claude Bernard Lyon 1,
UMR5208, Institut Camille Jordan, F-69622 Villeurbanne, France}

\author{Alexander G. Abanov}
\address{Department of Physics and Astronomy and Simons Center for Geometry and Physics, Stony Brook University, Stony Brook, NY 11794, USA}


\date{}                     


\begin{abstract}
We revisit the problem of computing the boundary density profile of a droplet of two-dimensional one-component plasma (2D OCP) with logarithmic interaction between particles in a confining harmonic potential. At a sufficiently low temperature but still in the liquid phase, the density exhibits oscillations as a function of the distance to the boundary of the droplet. We obtain the density profile numerically using Monte-Carlo simulations of the 2D OCP. We argue that the decay and period of those oscillations can be explained within a picture of the Wigner crystallization near the boundary, where the crystal is gradually melted with the increasing distance to the boundary. 
\end{abstract}


\tableofcontents

\section{Introduction}

The one-component plasma is a classical system of identical, charged particles, interacting via a repulsive potential. Here we focus on a two-dimensional plasma with a logarithmic potential placed in a uniform neutralizing background of opposite sign. We refer to this system as 2D OCP. The energy of such plasma is given by 
\begin{eqnarray}
    E &=& -q^2\sum_{1\leq j<k\leq N}\ln|z_j-z_k| + q^2\sum_{j=1}^N W(z_j,\bar{z}_j)\,.
 \label{EOCP}
\end{eqnarray}
Here $q$ is the charge of each particle and $N$ is the total number of particles. We use complex notations for particle coordinates $z=x+iy$, $\bar{z}=x-iy$. For a uniform neutralizing background corresponding to a one particle per unit area one has
\begin{eqnarray}
    W(z,\bar{z})=\frac{\pi}{2}  |z|^2 \,.
 \label{Wharm}
\end{eqnarray} 
All thermodynamic properties of the model are then determined by its partition function
\begin{eqnarray}
    \mbox{}\hspace{-0.4cm} 
    Z &=& \int \prod_{j=1}^N d^2z_j \, 
    \exp\left\{-\Gamma \left[-\sum_{1\leq j<k\leq N}\ln|z_j-z_k| 
    +  \sum_{j=1}^N  W(z_j,\bar{z}_j)\right]\right\}.
 \la{ZOCP} 
\end{eqnarray}
The model depends on two parameters: the dimensionless inverse temperature $\Gamma=q^2/k_BT$ and the number of particles $N$. In the following, we will be interested in the many-particle problem $N\gg 1$. The 2D OCP model (\ref{ZOCP}) is a classical model of statistical mechanics \cite{1980-BausHansen-PhysRep,caillol1982monte,deLeeuw} and it was extensively studied in the literature together with its various generalizations and extensions. It has resurfaced many times in connection to different problems. To mention a few: at  $\Gamma=2$ the Boltzmann weight is equal to the probability density for the exact many-electron wavefunction of the completely filled lowest Landau level, and Laughlin famously proposed that the analogy could be carried on to hold between a few states with fractional filling $\nu$ and the 2D OCP at $\Gamma=2/\nu$ \cite{laughlin1983anomalous}; at some special temperatures, the equilibrium density of particles of the 2D OCP is found to give the probability distribution function of eigenvalues of certain random matrix ensembles \cite{forrester2016analogies,forrester2010log}, like the complex Ginibre ensemble for $\Gamma=2$ \cite{ginibre1965statistical}; chiral matter, where a (constant-sign) vorticity patch in an incompressible inviscid two-dimensional fluid is understood as composed of discrete vortices of circulation proportional to $\Gamma$ \cite{wiegmann2013hydrodynamics,wiegmann2013anomalous}.

Despite an extensive literature, the phase diagram of the 2D OCP is still a subject of controversy (for a recent review see, e.g., \cite{Khrapak:2016ij}). It is believed that there is a melting phase transition at $\Gamma=\Gamma_m\approx 140$. The thermodynamic state is one of a crystal at low temperatures $\Gamma>\Gamma_m$ and a liquid (plasma) state at high temperatures $\Gamma<\Gamma_m$ \cite{alastuey1981classical,choquard1983cooperative}. As the most stable lattice at zero temperature is the triangular one \cite{Tkachenko:1966ux}, it seems reasonable to assume that this lattice persists until the melting temperature. The nature of the melting phase transition is less certain. One possibility is that the order is destroyed by a first-order phase transition. The other leading scenario is that the triangular crystal is melted by proliferating dislocations according to Berezinsky-Kosterlitz-Thouless-Halperin-Nelson-Young (BKTHNY) scenario  \cite{berezinsky1970destruction,berezinsky1972destruction,kosterlitz1973ordering,halperin1978theory,nelson1979dislocation,young1979melting} (see also Ref.~\cite{kleinert1989gauge} for review and further references). In the latter, dislocations of the triangular lattice decouple at the transition point, destroying the quasi-long-range translational order of the crystal. This scenario also predicts a possibility of an  intermediate hexatic phase so that, for some $\Gamma<\Gamma_m$, the system's translational order is broken but there is still a remaining quasi-long-range orientational order, which is then destroyed at some higher temperature $\Gamma_m'<\Gamma_m$. \cite{nelson1979dislocation} While the nature of the phase transition is non-universal and depends on the interparticle potential \cite{kleinert1989gauge,kapfer2015two,khrapak2018note}, for the logarithmic interactions considered in this work most numerical studies see a single weakly first-order phase transition at $\Gamma=\Gamma_m\approx 140$ with no compelling evidence for the intermediate hexatic phase \cite{caillol1982monte,deLeeuw,choquard1983cooperative,radloff1984freezing}. In the following, we assume that there is a single melting transition at $\Gamma_m$. Regardless of the nature of this transition, it is clear that the softening of the crystal by dislocation pairs, essential for the BKTHNY theory, is important on the crystal side of the transition (low temperatures). It might help explain, for example, why the transition occurs at a temperature which is significantly lower than the typical temperature scale $\Gamma\sim 1$.

While it seems that, numerically, nothing dramatic happens at temperatures above melting $\Gamma<\Gamma_m\approx 140$, an important change occurs in the boundary density profile at exactly $\Gamma=2$. As we mentioned, $\Gamma=2$ is the special point where the Boltzmann weights of (\ref{ZOCP}) can be understood through Laughlin's plasma analogy as a probability distribution of non-interacting electrons completely filling the lowest Landau level in a constant magnetic field. The problem of free fermions in a magnetic field can then be solved exactly. In particular, one can compute analytically the density $\rho(r)$ \cite{jancovici1981exact,jancovici1982classical}, which is a function of the radius $r$. It is equal to $1$ in the bulk of the droplet and decays monotonously to zero at the boundary (see Figure \ref{fig:beta1}). For $\Gamma>2$, however, the character of the density profile changes from monotonic ($\Gamma<2$) to oscillating \cite{badiali1983surface,datta1996edge,morf1986monte,can2014edgelaughlin}, analogously to the character of the bulk pair correlation function \cite{caillol1982monte,levesque2000charge,morf1986monte}. This transition in the behavior of the boundary density profile has been studied analytically using the expansion around $\Gamma=2$ \cite{jancovici1981exact,can2014edgelaughlin,can2015exact} and by a combination of analytical and numerical methods available in the theory of simple fluids such as hypernetted-chain approximation \cite{datta1996edge}, molecular dynamics \cite{deLeeuw,choquard1983cooperative} and Monte-Carlo \cite{morf1986monte}. Although this has been repeatedly verified by numerical calculations, a full analytical understanding of this phenomenon is lacking. A few works clarified the appearance of an overshoot singularity, which can be seen as the first peak of the oscillations, in the density \cite{zabrodin2006large,can2014edgelaughlin} and pair correlation function \cite{jancovici1981exact}, but the properties of the oscillations themselves appear as non-perturbative effects which could not yet be captured by the large-N and temperature expansions employed in there.

In this work, we will approach the problem from a different starting point. We assume that, at sufficiently low temperatures but still above the freezing temperature, $2\ll \Gamma <\Gamma_m$, a good starting point is to think that the OCP freezes near the boundary of the droplet. Then the oscillations of the density near the boundary correspond to the crystal planes (actually, lines for the 2D crystal) smeared by thermal fluctuations. These oscillations decay into the bulk of the droplet, interpolating into the bulk liquid phase. We will show that this picture is consistent with new Monte-Carlo data obtained in the temperature range $2\ll \Gamma <\Gamma_m$ and can be extended almost all the way to $\Gamma=2$. 

In the conclusion of our brief review, we mention that there is significant literature on the mathematics of the 2D OCP. In the mathematical physics community, many works are studying the effects of the background curvature and topology on 2D OCP defined on curved surfaces \cite{zabrodin2006large,ferrari2014fqhe,can2014fractional}. On the mathematical side, the main contributions lie at the interface between analysis and probability theory. A typical result concerns the large $N$ behavior of $ \sum_{i=1}^N f(z_i)$ for $f$ a smooth enough function, supported on scales much larger than inter-particle distance. Its mean is given by a simple ``liquid droplet'' continuous description like the one discussed in section~\ref{sec:2DOCP}, and the fluctuations are given by a free field theory  \cite{RiderVirag,LebleSerfaty,BBNY}. Such Gaussian fluctuations are of order one, and inversely proportional to $\Gamma$, for all $\Gamma>0$. Hence the oscillations we are studying in this paper are nowhere to be seen. There is no contradiction, however, since these oscillations occur precisely on scales of the order of the inter-particle distance. Hence they are averaged out and disappear in the regime where rigorous techniques can be used. This is also what makes crystallization challenging to study from a mathematical perspective. In fact, even the widely accepted fact that the triangular lattice minimizes the Coulomb energy at infinite $\Gamma$ is still a conjecture \cite{CohnKumar}. Note also that the smoothness assumption on $f$ is important; for example, full counting statistics in a spatial region 
$A$, which corresponds to $f$ being an indicator function, leads to different scaling behaviors, with fluctuations of the order of the boundary length of $A$ \cite{Charles2020,EstienneStephan}. 

We start with a brief introduction into the scales and properties of the OCP in the high-temperature phase $\Gamma\leq 2$ in Section~\ref{sec:2DOCP} to establish notations and set up the problem. In Section~\ref{sec:MC} we present results of new  Monte-Carlo simulations. We analyze numerically the oscillations of density near the edge of the droplet. In Section~\ref{sec:freezing} we argue that at sufficiently low temperatures $\Gamma\sim 120$ the boundary density profile suggests the picture of the plasma being frozen into a triangular crystal near the edge of the droplet. We compute the density of the triangular crystal near the boundary taking into account the smearing of lattice planes by thermal fluctuations. We find in Section~\ref{sec:freezing} that the obtained density profiles are quantitatively consistent with the results of our Monte-Carlo simulations. We summarize and discuss possible future directions of research in Section~\ref{sec:conclusions}. Some details of computations are relegated to appendices.

\section{Statistical mechanics of 2D OCP}
 \label{sec:2DOCP}

The statistical mechanics model of the 2D OCP in a constant background of opposite charge is defined by the partition function (\ref{ZOCP}). We will be interested in the corresponding density profile, given by the thermodynamic average of the microscopic density of particles
\begin{eqnarray}
     \rho(z) =\left\langle \sum_{i=1}^N\delta^{(2)}(z-z_i)\right\rangle_\Gamma\,,
 \label{eq:density}
\end{eqnarray}
where the subscript $\Gamma$ indicates that the averaging over particles' coordinates $z_i$ is done at the corresponding inverse temperature. The angular brackets in (\ref{eq:density}) denote the thermodynamic average with respect to (\ref{ZOCP}). In (\ref{eq:density}) and below we use an abbreviated notation $\rho(z)$ instead of the full notation $\rho(z,\bar{z})$. 

As we are interested in the case $N\gg 1$ it is tempting to approximate the statistical ensemble of the system in terms of coarse-grained variables, averaging over a fluctuating smooth density of particles $\rho(z)$ instead of the particles' microscopic coordinates $z_i$, $i=1,\ldots,N$. We refer the reader to Ref.~\cite{zabrodin2006large} for a systematic treatment which realizes this through the $1/N$ expansion and further references within. As a result the exact partition function (\ref{ZOCP}) is replaced by a functional integral over the density field $\rho(z)$,
\begin{eqnarray}
    Z = \int D\rho \; \exp\left\{-\Gamma F[\rho]\right\}\,,
 \label{Zrho}
\end{eqnarray}
where
\begin{eqnarray}
    F[\rho] &=& -\frac{1}{2}\int d^2z\,d^2z'\, \rho(z)\ln|z-z'|\rho(z') 
    +\int d^2z\, \rho(z) W(z)
 \nonumber \\
    &-& \left(\frac{1}{4}-\frac{1}{\Gamma}\right) \int d^2z\,\rho(z)\ln\rho(z)\,.
 \la{E-Dyson}
\end{eqnarray}
The two terms in the first line of (\ref{E-Dyson}) are obtained by a naive replacement of the summation in (\ref{EOCP}) by integration with density. The origin of the second line is more subtle. The part with coefficient $1/4$ originates from the short distance regularization of the logarithmic potential while the part with $1/\Gamma$ comes from the change of the measure in the partition function from $\prod d^2z_i$ to $D\rho$. It is also important to remember that the functional integral in (\ref{Zrho}) should be taken over semi-positive-definite fields $\rho(z)\geq 0$ constrained by a fixed total number of particles 
\begin{eqnarray}
    \int d^2z\,\rho(z)=N \,.
 \label{eq:norm}
\end{eqnarray} 

Taking the variation of (\ref{E-Dyson}) with respect to $\rho(z)$ and then applying the Laplacian $\Delta$ to the result we obtain the following saddle-point equation
\begin{eqnarray}
    \rho-\bar{\rho} + \frac{\Gamma-4}{8\pi\Gamma}\Delta\ln\rho =0\,.
 \label{MFE}
\end{eqnarray}
Here we defined the background charge density $\bar{\rho}$ as
\begin{eqnarray}
    \bar{\rho} \equiv  \frac{\Delta W}{2\pi}=1 \,,
\end{eqnarray}
where we used the choice of the potential (\ref{Wharm}). Note that this is equivalent to a choice of the length unit, which we adopt along the paper except where explicitly stated otherwise.

The saddle-point equation (\ref{MFE}) is known as the Mean Field Equation (MFE) and is the subject of a number of studies \cite{lin2000uniqueness,lin2006uniqueness,ricciardi1998periodic}. Its obvious solution is
\begin{eqnarray}
    \rho(z)=1 \,,
\end{eqnarray}
corresponding to the exact screening of the background charge by the OCP particles. In a finite system, however, the solution is more complicated. Indeed, while the solution $\rho(z)=1$ is the correct minimum of $F[\rho]$ in the bulk, the correct minimum outside of the domain where the particles are concentrated is $\rho(z)=0$.\footnote{The vanishing solution $\rho(z)=0$ does not come from (\ref{MFE}) but from the boundary of the domain of functional integration. Remember that the functional integral should be taken over the fields satisfying $\rho(z)\geq 0$.} This means that as a good initial guess one can take the circular ``droplet'' 
\begin{eqnarray}
    \rho_0(r) = \theta(R-r)\,,\qquad R=\sqrt{\frac{N}{\pi}}\,,
 \label{rho0r}
\end{eqnarray}
where $\theta(x)$ is a step function, $r=|z|$ and the radius of the droplet $R$ is found from the normalization condition (\ref{eq:norm}). While this solution describes the main effect of screening, it does not take into account $1/N$ corrections represented by the last term of MFE (\ref{MFE}). These corrections are generally singular in the $1/N$ expansion \cite{zabrodin2006large} and become smooth only when terms nonperturbative in $1/N$ are taken into account. The latter non-perturbative effects are the subject of this work.

\begin{figure}[H]
  \centering
    \includegraphics[width=0.6\textwidth]{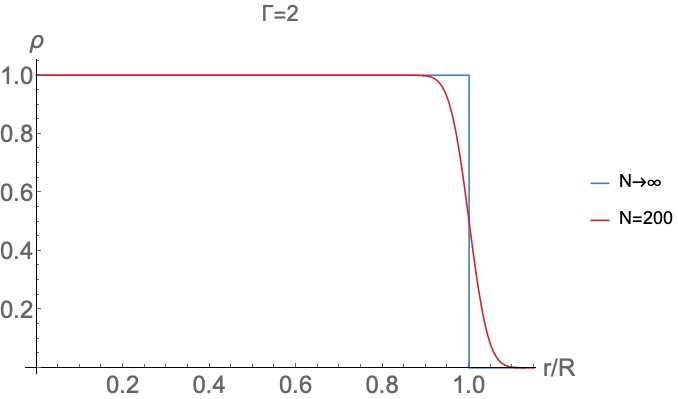}
  \caption{Exact density profile at $\Gamma=2$ for $N=200$ particles. The infinite $N$ result is also shown to illustrate the non-perturbative effect of smearing, present at any finite $N$. In plots of this form we present $r$ and $\rho$ in units of the radius $R=\sqrt{N/\pi}$ and the bulk density $\bar{\rho}$, respectively.}
  \label{fig:beta1}
\end{figure}

As a first example of such non-perturbative effects consider the case of the special temperature of $\Gamma=2$ at which the density can be exactly computed from the original partition function (\ref{ZOCP}) by mapping to free fermions \cite{jancovici1981exact,jancovici1982classical}. The result of the computation is (see figure \ref{fig:beta1}) 
\begin{eqnarray}
    \rho^{\Gamma=2}(r)=\frac{\Gamma(N,\pi r^2)}{\Gamma(N)}\, ,
 \label{eq:erfc}
\end{eqnarray}
where $\Gamma(s,x)$ and $\Gamma(s)$ in the right-hand side denote the incomplete Gamma function, and the Gamma function respectively: 
\begin{eqnarray}
    \Gamma(s,x) = \int_{x}^{\infty}e^{-t}t^{s-1}dt, \,\,\,\,  \Gamma(s)=\int_{0}^{\infty}e^{-t}t^{s-1}dt,
\end{eqnarray}
and should not to be confused with the inverse temperature temperature $\Gamma=2$.
We express the density (\ref{eq:erfc}) as a function of $\xi\equiv \frac{r^2-R^2}{2R^2}$. Close to the boundary of the droplet, $r\approx R$, the variable $\xi\approx (r-R)/R$ coincides with the rescaled distance to the boundary. After taking the Fourier transform of (\ref{eq:erfc}) and performing straightforward computations (see \ref{app:G2boundary}) we find 
\begin{eqnarray}
    \rho^{\Gamma=2}(\xi)=\theta(-\xi) +\int_{-\infty}^{+\infty}\frac{dk}{2\pi}\, \frac{e^{ik\xi}}{ik} \left(1-e^{i\frac{k}{2}} \left(1+\frac{ik}{2N}\right)^{-N}\right)\,.
\end{eqnarray}
Expanding in $1/N$ we obtain (c.f. \cite{zabrodin2006large}) a singular expansion of the density near the droplet boundary
\begin{eqnarray}
    \rho^{\Gamma=2}(\xi) &=& \theta(-\xi) -\frac{1}{8N}\delta'(\xi)
    +\frac{1}{24N^2}\delta''(\xi)
 \nonumber \\
 	&-& \frac{1}{128 N^2}\left(1+\frac{2}{N}\right)\delta'''(\xi)+\ldots\,.
 \label{exp10}
\end{eqnarray}
The series (\ref{exp10}) is asymptotic in $1/N$. The first few terms of (\ref{exp10}) do not capture effects non-perturbative in $1/N$ such as the smearing of the boundary profile at $r-R\sim 1$  (or $\xi\sim 1/\sqrt{N}$) as well as the exponentially small corrections to the density far away from the boundary at $|r-R|\gg 1$. These effects are present in the exact expression (\ref{eq:erfc}) but are lost when only the first few terms of (\ref{exp10}) are considered.


In this work we are going to focus on other effects which are non-perturbative in $1/N$. Namely, we will see that at $\Gamma>2$ the density profile develops oscillations with period of the order of interparticle distance. The amplitude of these oscillations decay exponentially into the bulk of the droplet. Both the existence of oscillations and the nature of their decay is non-perturbative in $1/N$.

\begin{figure}[H]
    \centering
    \begin{tabular}{c c}
      \includegraphics[width=0.395\textwidth]{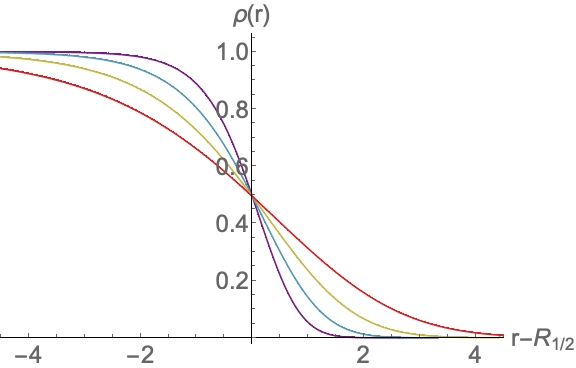}
         &
         \includegraphics[width=0.505\textwidth]{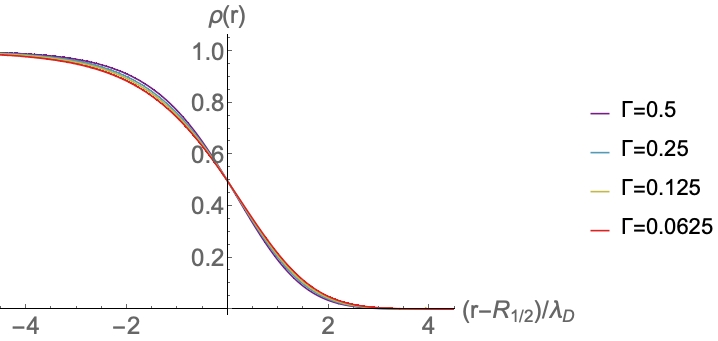}
    \end{tabular}
     \caption{In the Debye-H\"uckel regime of $\Gamma\ll 1$, the boundary density profile has no oscillations, like in the $\Gamma=2$ case, and the smearing at the boundary is controlled by the Debye length $\lambda_D=(2\pi\Gamma)^{-1/2}$. In the left, the density is plotted for different values of $\Gamma<1$, $N=400$ particles. The horizontal axis shows the distance from the classical radius of the droplet, which is defined here as the position at which $\rho=1/2$. In the right, the same curves are shown, but the radial coordinate is rescaled by the Debye length, so that the curves collapse on top of one another.}
    \label{fig:dhpic}
\end{figure}

Before going to the regime of interest for this paper, $2<\Gamma<140$, let us briefly consider high temperatures  $\Gamma\ll 2$. For small $\Gamma$, the Debye-H\"uckel approximation is applicable \cite{caillol1982monte,jancovici1982classical,vsamaj2004two}, and the density at the boundary remains a monotonously decreasing function of the radius like in the $\Gamma=2$ case, but with a smearing which scales with the Debye length $\lambda_D=(2\pi\Gamma)^{-1/2}$, as shown in figure (\ref{fig:dhpic}).

In the regime of large $\Gamma$ the Debye length is formally smaller than the inter-particle distance and both the mean-field and the Debye-H\"uckel approximations are not self-consistent at the boundary. Thus alternative approximations should be used.

\section{Intermediate temperatures \texorpdfstring{$2\ll \Gamma<\Gamma_m\approx 140$}{2<<G<<Gm=140}: Monte Carlo Results}
 \label{sec:MC}

As the freezing of 2D OCP occurs at a numerically large value of $\Gamma\approx 140$ \cite{choquard1983cooperative,Khrapak:2016ij, alastuey1981classical} we can investigate the regime of $\Gamma$ large but still below freezing $2\ll \Gamma<\Gamma_m\approx 140$. In this section, we investigate this regime using Monte Carlo simulations and observe that density oscillations develop near the boundary of the droplet. For now, we focus on presenting numerical observations leaving possible interpretations for the following sections.

All data shown in this paper are obtained using a simple update scheme that we now explain. Given a configuration of the $N$ particles, we pick one particle at random and then propose a move elsewhere using a symmetric probability distribution. The move is then accepted or rejected according to the Metropolis-Hastings criterion. Due to the specific form of the energy (\ref{EOCP}), this can be achieved in $O(N)$ time. We found this to outperform more sophisticated but slower schemes. The chosen probability distribution also favors moves to a distance $O(1)$ to get a significant acceptance rate, but longer moves are also possible. All random numbers are obtained using the 64-bit Mersenne twister generator.

A given MC run is initialized by picking random uniform points for the positions of the particles, in the disk of radius $R$. We then try many updates, make measurements every fixed number of updates, after a given burn-in time. We found that these parameters have to be chosen carefully when $\Gamma>140$ and $N>1000$. In that case, we chose about $50N$ MC tries between measures, performed about $4\times 10^5$ measures, and took a burn-in time as high as one quarter of the total simulation time. 

To make sure that proper thermalization is achieved, we make several independent runs (in practice about 12, and double that for $\Gamma>140$), compare the final results, and check that those are very close. Finally, we average over all the runs, to further improve statistics. This final average is the data shown in all subsequent plots. At the free fermion point, we also checked that this procedure reproduces exact finite $N$ results for the density, to high precision. Another check is presented in the next subsection.

\subsection{Sum rules}
\label{sec:sumrules}

\begin{figure}[H]
\centering
\includegraphics[width=0.7\textwidth]{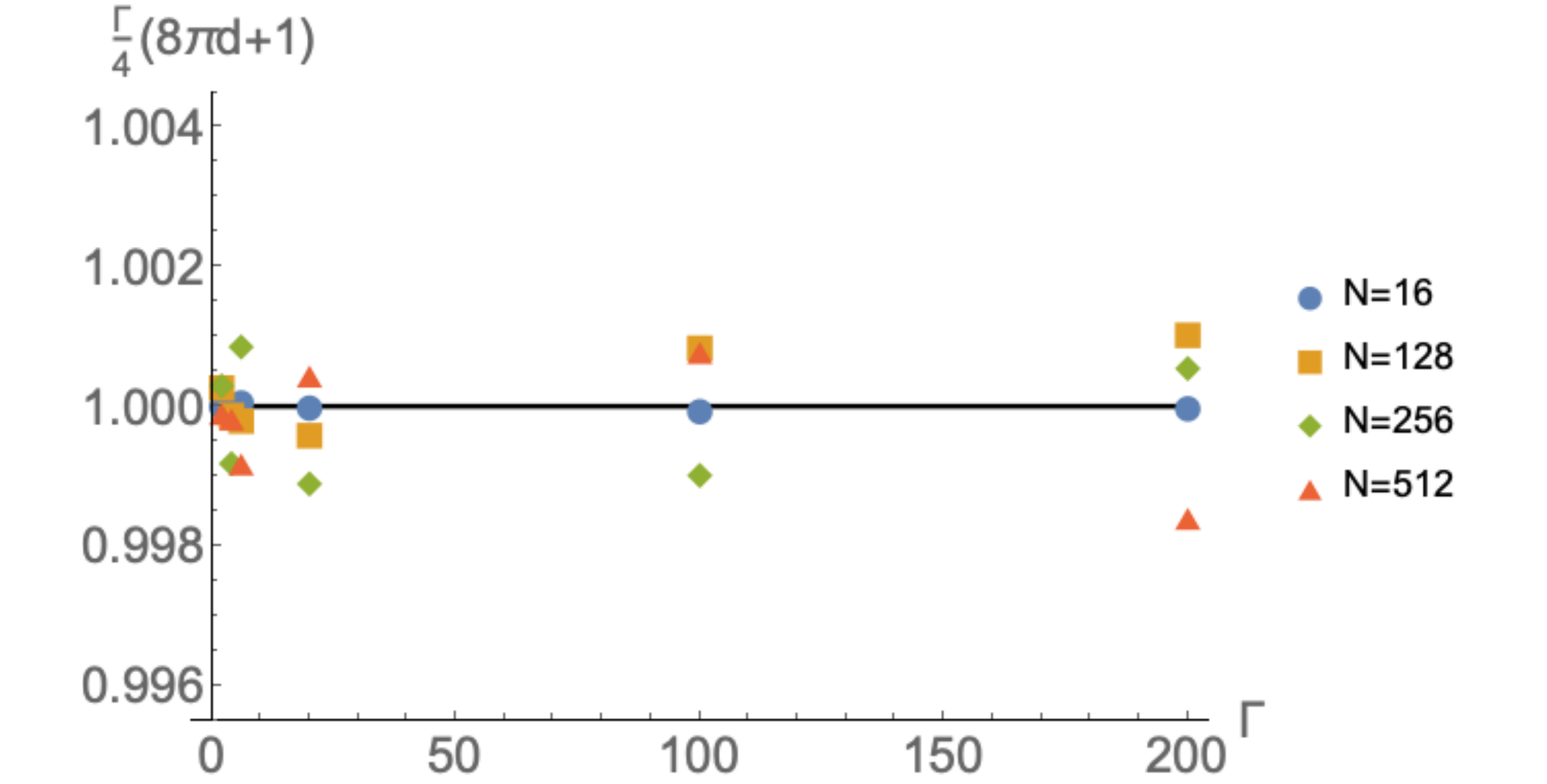}
\caption{Numerical check of the sum rule as function of $\Gamma$ (the values $\Gamma=2,4,6,20,100,200$ are shown) for $N=16,128,256,512$. We plot the Monte Carlo estimate of $\frac{\Gamma}{4}\left(8\pi d+1\right)$ which should equal $1$. As can be seen the error is particularly small, less $0.2\%$. Data for a smaller number of particles with considerably better statistics is also shown.}
\label{fig:sumrules}
\end{figure}

The 2D OCP studied here satisfies a number of sum rule identities \cite{martin1988sum,kalinay2000sixth,zabrodin2006large}. The first one is simply the statement that the total number of particles is $N$. The second one is non trivial, and serves as a strong check of our numerical procedure. To introduce it let us define the quantity
\begin{equation}
    d=\frac{1}{4N}\int d^2 x (\rho-\rho_0) r^2\,,
 \label{ddef}
\end{equation}
which, in the large $N$ limit, is related to the boundary dipole density. $d$ can be computed exactly for finite $N$ (see \ref{app:sumrules}):
\begin{equation}
    d=-\frac{1}{8\pi}\left(1-\frac{4}{\Gamma}\right)\,,
\end{equation}
which can be rewritten as $\frac{\Gamma}{4}\left(8\pi d+1\right)=1$. A numerical check of the latter identity is presented in figure \ref{fig:sumrules}, and shows impressive agreement. This agreement within 0.2\% is a testament to the accuracy of our Monte Carlo simulations and is an evidence that the system thermalizes well even at low temperatures (large $\Gamma$).

\subsection{Oscillations near the edge}
\label{sec:oscillations}

\begin{figure}[H]
  \centering
    \includegraphics[width=0.7\textwidth]{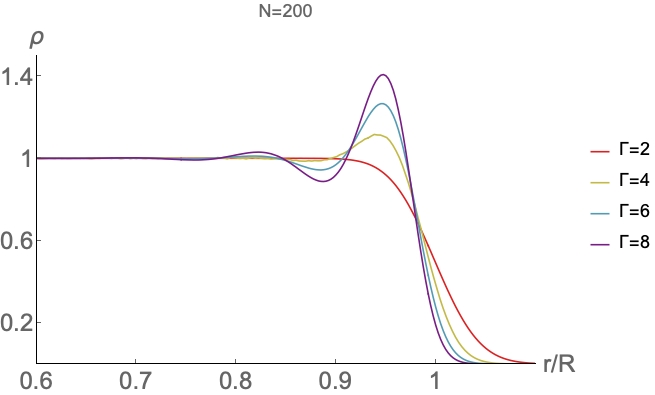}
  \caption{Appearance of oscillations in the density profile as $\Gamma$ is increased from $\Gamma=2$ to $\Gamma=8$. These are simulations for $N=200$ particles.}
  \label{fig:lowgamma}
\end{figure}

For $\Gamma>2$, oscillations appear close to the edge \cite{badiali1983surface,datta1996edge,morf1986monte,can2014edgelaughlin}, as one can see in figure \ref{fig:lowgamma}. These oscillations are at the scale of the inter-particle distance and appear also in the pair correlation function of the density \cite{jancovici1981exact,levesque2000charge}. In the large $N$ limit, in which the boundary becomes sharp when plotted as a function of $r/R$, the first peak develops into an overshoot singularity \cite{zabrodin2006large}.

\begin{figure}[H]
  \centering
    \includegraphics[width=0.7\textwidth]{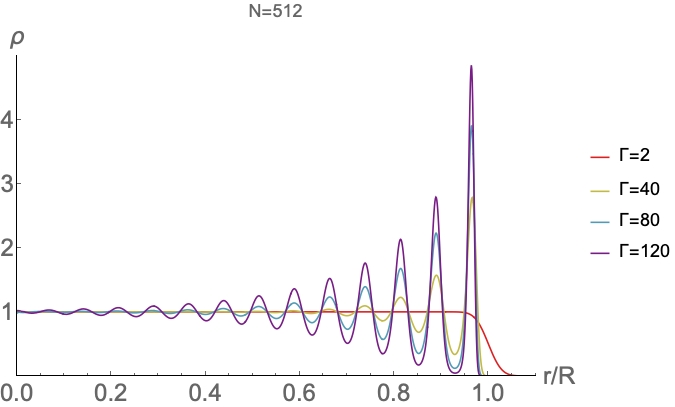}
  \caption{Differences in density profile as $\Gamma$ is increased from $\Gamma=2$ to $\Gamma=120$. Here, $\Gamma=2$ is a plot of the exact analytical expression while the other curves are simulations for $N=512$ particles.}
  \label{fig:increasebeta}
\end{figure}

We also observe already in figure \ref{fig:lowgamma} that the number of visible oscillations increases with the increasing value of $\Gamma$. This trend continues as can be seen from the following numerical results, probing this phenomenon at much smaller temperatures. A few results are summarized in figure \ref{fig:increasebeta}, which gives the radial density profile for values of $\Gamma$ ranging from $\Gamma=2$ to $\Gamma=120$, for $N=512$ particles. In this whole range, the oscillations start from the edge and are damped into the bulk, while the number of visible oscillations increases with $\Gamma$ starting from $\Gamma>2$.

\begin{figure}[H]
  \centering
    \includegraphics[width=\textwidth]{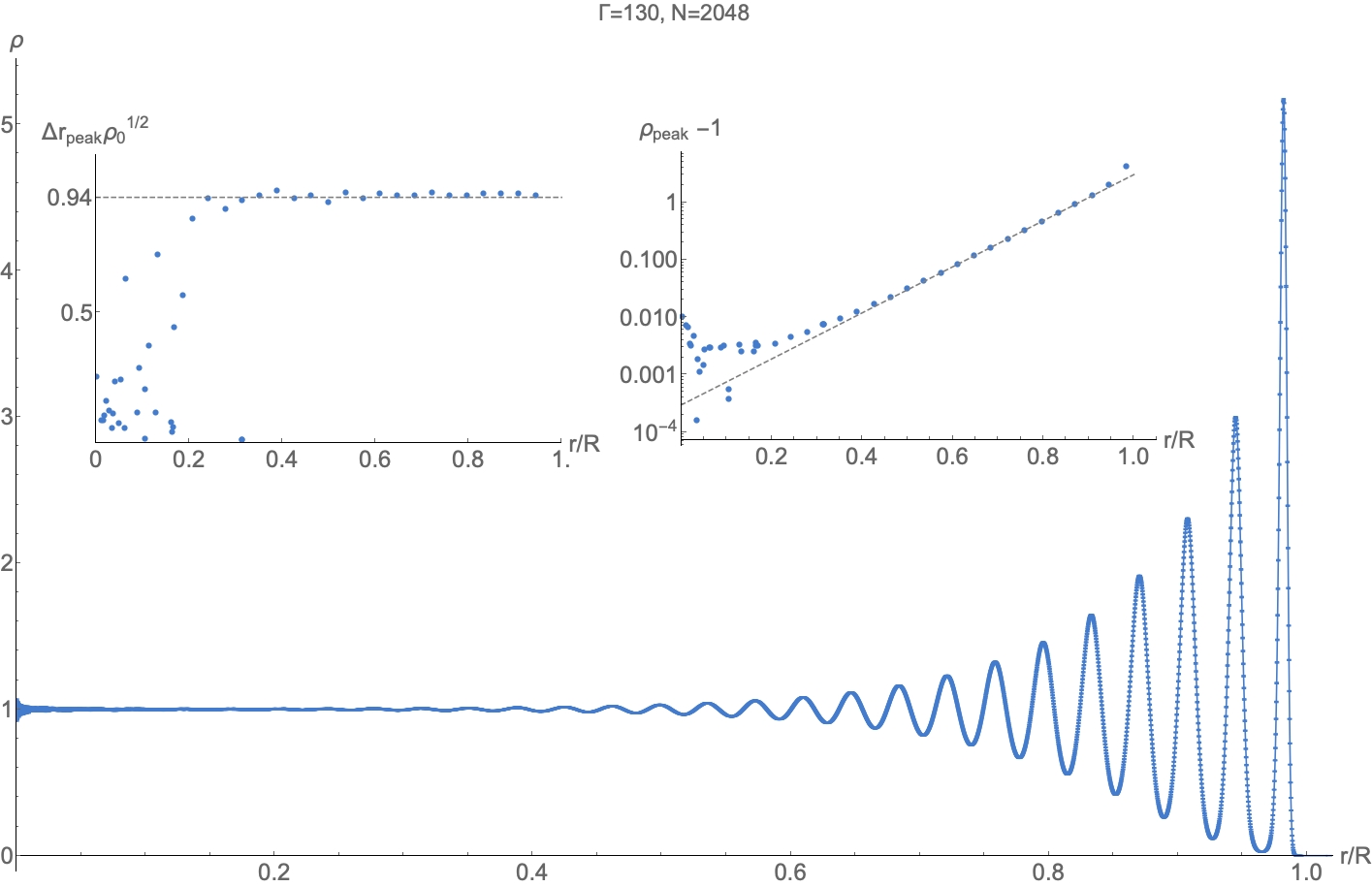}
  \caption{Density profile of the 2D OCP at $\Gamma=130$, $N=2048$. The points have (almost invisible) error bars to show the high precision of the data. The solid line is just an interpolation between the points. The left inset gives the distance between consecutive peaks as a function of the radius and shows that this is a constant equal to $0.94\pm 0.02$ close to the edge of the droplet. The right inset presents the peaks in log-linear scale demonstrating that the decay of oscillations is exponential. }
  \label{fig:gamma130}
\end{figure}

\begin{figure}[H]
    \centering
    \begin{tabular}{c c}
      \includegraphics[width=0.45\textwidth]{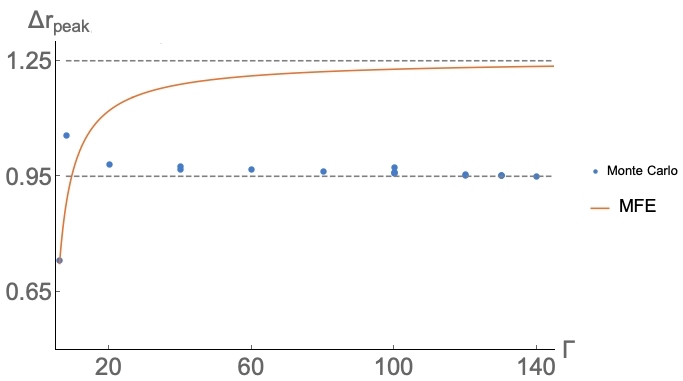}
         &
         \includegraphics[width=0.45\textwidth]{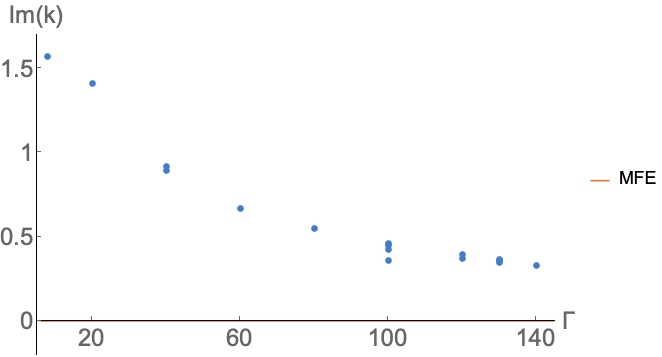}
    \end{tabular}
     \caption{Left: Comparison of the distance between peaks in the oscillations observed in simulations of the 2D OCP at various inverse tempeartures $\Gamma$ (points) with the prediction from linear waves in the mean-field theory (\ref{eq:kgamma}) (curve). There is only a possibility of an agreement at large temperatures (small $\Gamma$). Right: Imaginary part of the wave vector of oscillations in density profiles found from Monte Carlo simulations at different values of $\Gamma$ shows that the damping increases with temperature. The mean-field equation does not predict any damping of these oscillations.}
    \label{fig:peaks}
\end{figure}

For a given temperature and number of particles, the peaks are roughly equally spaced. This can be seen clearly from figure \ref{fig:gamma130} at $\Gamma=130$. The inter-peak distance is of the order of the inter-particle distance  $\Delta r_{peak} = 0.94\pm 0.02$ and is independent of $\Gamma$ and $N$. This can be contrasted with the linear waves coming from the mean-field theory. Linearization of the MFE (\ref{MFE}) around the constant solution, $\rho=1+\delta\rho\, e^{ikr}$ gives a wave vector and the wavelength of oscillations
\begin{equation}
 \label{eq:kgamma}
    k(\Gamma)=\sqrt{\frac{8\pi\Gamma}{\Gamma-4}}\,,
    \qquad
    \lambda =\frac{2\pi}{k}= \sqrt{\frac{\pi}{2}\left(1-\frac{4}{\Gamma}\right)}\,.
\end{equation}
First of all, we observe that the corresponding wavelengths of these waves disagree with the inter-peak distances observed in our simulations except perhaps for small values of $\Gamma$, as shown in figure \ref{fig:peaks}. Indeed, at large $\Gamma$ we have $\lambda \approx \sqrt{\pi/2}\approx 1.25$, significantly different from the numeric result $0.94\pm 0.02$. 

A second important observation about the oscillations is that they are exponentially damped into the bulk as can be seen in the right inset in figure \ref{fig:gamma130}. This aspect is absent in the mean-field theory (\ref{MFE}), which leads to real (i.e., no damping) wave vectors (\ref{eq:kgamma}). The damping is independent of $N$ and decreases with $\Gamma$ as shown in figure \ref{fig:peaks}, leading to more visible peaks at lower temperatures.

Let us also mention that the positions of the peaks, measured from the boundary, are independent of $\Gamma$ (figure \ref{fig:increasebeta}) and depend on $N$ only through the length-scale set by the bulk density $\bar{\rho}^{1/2}$. On the other hand, the number of visible oscillations is only a function of $\Gamma$.

\subsection{Squeezing}
\label{sec:squeezing}

\begin{figure}[H]
  \centering
    \includegraphics[width=0.7\textwidth]{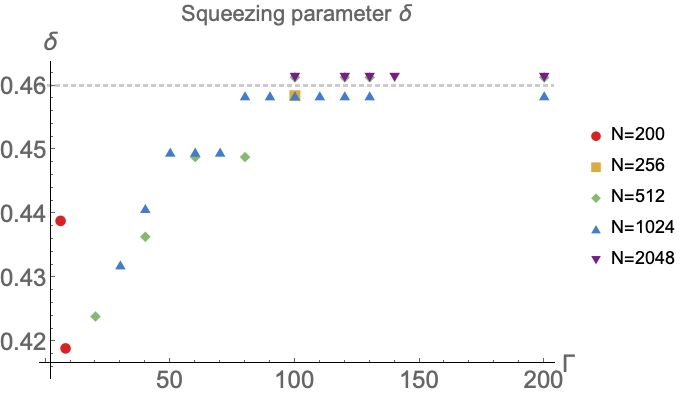}
  \caption{The squeezing of the droplet, defined as the distance $\delta$ between the position of the last peak of the density profile $\rho(r)$ and the classical radius, is seen to saturate at some value slightly above $0.46$.}
  \label{fig:squeezing}
\end{figure}

Another non-perturbative effect known to be present in the density profile of the 2D OCP is the squeezing of the droplet \cite{bogatskiy2019edge}. Here we adopt a simple definition of the squeezing (different from the one used in \cite{bogatskiy2019edge}) as the distance $\delta$ between the last peak of the density profile $\rho(r)$ and the classical radius of the droplet $R=\sqrt{N/\pi}$. This definition is applicable only for $\Gamma>2$ as the density profile does not have any peaks for $\Gamma\leq 2$.  With this definition, the squeezing is seen to be roughly independent of $N$ and increasing with $\Gamma$, saturating at a value slightly above $\delta\approx 0.46$ in units corresponding to $\bar\rho=1$, as shown in figure \ref{fig:squeezing}.

In the following section, we propose a simple physical picture that naturally explains these numerically observed effects: freezing at the edge.

From the perspective of the $1/N$ expansion and the mean-field approach, which seems to be a good approximation for the fluid bulk of the 2D OCP in the range $2<\Gamma<\Gamma_m\approx 140$, the density oscillations are nonperturbative. Therefore, in this work, we take a different ``low temperature'' starting point -- the triangular crystal formed by charges.

\section{Freezing at the edge}
 \label{sec:freezing}

Let us assume that the 2D OCP is frozen at the boundary and forms a crystalline triangular lattice. Then the peaks of density oscillations simply label the positions of crystal planes. As the bulk phase at $\Gamma<\Gamma_m$ is fluid, the crystal is melted by thermal fluctuations far away from the boundary which explains the decay of density oscillations.

\subsection{Oscillations and Squeezing}

\begin{figure}[H]
    \centering
    \begin{tabular}{c c}
      \includegraphics[width=0.33\textwidth]{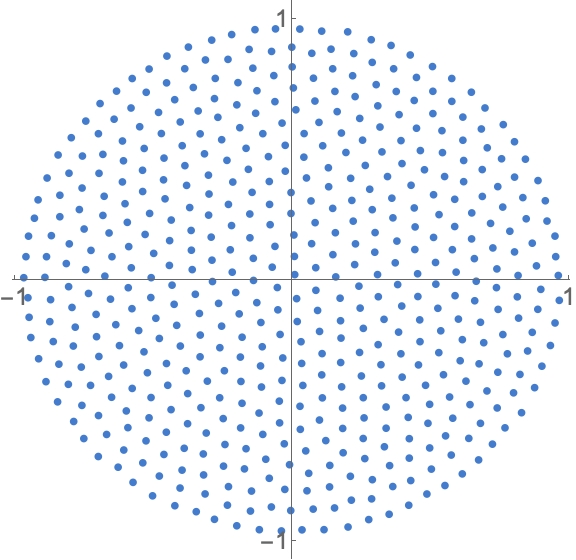}
         &
         \includegraphics[width=0.57\textwidth]{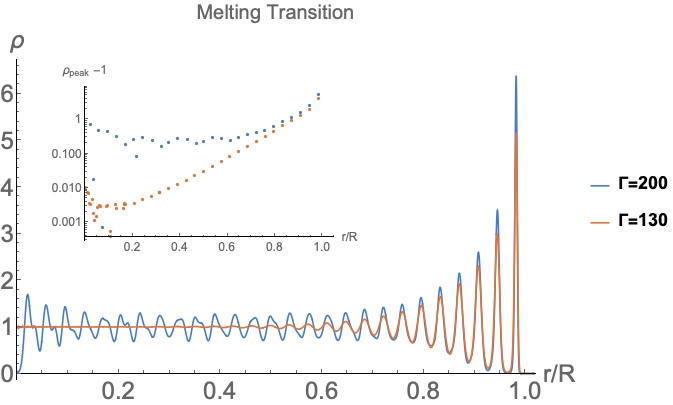}
    \end{tabular}
    \caption{At $\Gamma=200$, the oscillations in the radial density persist into the bulk, which can be contrasted to the exponential decay seen at $\Gamma=130$ the plots are for $N=2048$. In the left, a snapshot at $\Gamma=200$ for $N=512$ particles shows the triangular lattice structure.}
    \label{fig:crystal}
\end{figure}

To support this point of view let us compare density profiles obtained numerically for temperatures below and above the bulk melting transition $\Gamma_m\approx 140$. At $\Gamma>\Gamma_m$, the triangular crystal is the phase of the bulk. In the density profile, the difference between the two phases can be seen from the fact that in the crystal ($\Gamma>\Gamma_m$) the oscillations do not decay exponentially as they do in the fluid phase ($\Gamma<\Gamma_m$), as shown in figure \ref{fig:crystal}. Notice, however, that close to the boundary the profiles are close to each other and have the same period of oscillations.

\begin{figure}[H]
\centering
\begin{tikzpicture}[scale=1.4]
\pgfmathsetmacro\sq{0.5*sqrt(3)}
\draw [dashed] ({2*\sq},2) -- ({2*\sq,3.1});
\draw [dashed] ({3*\sq},1.5) -- ({3*\sq,3.1});
\draw [dashed] ({3.5*\sq},1.25) -- ({3.5*\sq,3.1});
\foreach \y in {0,1,2,3}{
\draw (0,\y) -- ({3*\sq},{-3/2+\y});
}
\draw (0,0) -- ({3*\sq},{3/2});
\foreach \y in {1,2}{
\draw (0,\y) -- ({(3-\y)*\sq},{(3-\y)/2+\y});
\draw ({\y*\sq},{-\y/2}) -- ({3*\sq},{3/2-\y});
}
\foreach \x in {0,1,2,3}{
\draw ({\x*\sq},-\x/2) -- ({\x*\sq},3-\x/2);
\foreach \y in {0,1,2,3}{
\filldraw[red] ({\x*\sq},\y-0.5*\x) circle (0.08cm);
}
}
\draw[blue,ultra thick] ({2.5*\sq},-0.25) -- ({3.5*\sq},0.25) -- ({3.5*\sq},1.25) -- ({2.5*\sq},0.75) -- cycle;
\filldraw[blue!40!white,opacity=0.4] (0,0) -- ({3.5*sqrt(3)/2},-1.75) -- ({3.5*sqrt(3)/2},1.25) -- (0,3) -- cycle;
\draw[<->,thick] ({3*\sq},2) -- ({3.5*\sq},2);
\draw ({3.25*\sq},2.25) node {$\delta$};
\draw[<->,thick] ({2*\sq},2.5) -- ({3*\sq},2.5);
\draw ({2.5*\sq},2.75) node {$h$};
\end{tikzpicture}

    \caption{A segment of a triangular lattice with lattice constant $a$ is shown. The blue rhombus is a convenient choice of the unit cell of the lattice. The area of the unit cell is given by $\frac{\sqrt{3}}{2}a^2$. The interplane spacing is given by $h=a\frac{\sqrt{3}}{2}$. The region shaded in blue represents the uniform positive background. The last crystal line at the edge of the droplet is the distance $\delta=h/2$ from the outer edge of the background (classical radius). }
    \label{fig:trian-lattice}
\end{figure}
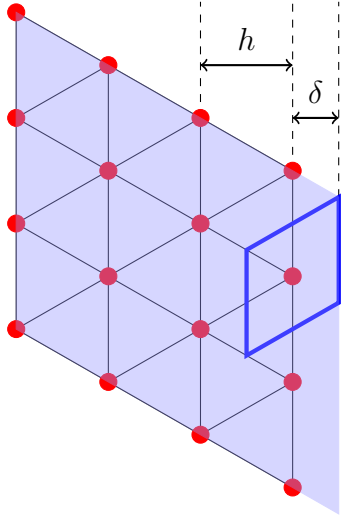

The wavelength of oscillations found numerically in the previous section $\Delta r=0.94\pm 0.02$ is compatible with the distance $h$ between lattice planes (in this case, lines) of a triangular lattice. Indeed, as it is illustrated in Figure~\ref{fig:trian-lattice}, the area per particle is given by $\frac{\sqrt{3}}{2}a^2$, where $a$ is the lattice constant and
\begin{equation}
    \bar{\rho}=\frac{2}{a^2\sqrt{3}}\,,\; h=a\frac{\sqrt{3}}{2}
    \quad \Rightarrow\quad  h\sqrt{\bar{\rho}}=\frac{3^{1/4}}{2^{1/2}}\approx 0.93\,,
\end{equation}
where we briefly restored explicitly our length unit, the bulk density $\bar{\rho}$, for clarity.

The squeezing can likewise be explained. Note that, in finding the distance between lattice planes above, we used the charge screening to determine the size of the lattice constant. We now consider the charge screening at the edge of a triangular crystal, which we take to occupy a half-plane, for simplicity. It is convenient to think of a triangular lattice as obtained by translations of a unit cell given by a rhombus surrounding a vertex of the lattice (see Figure~\ref{fig:trian-lattice}). The charge screening here means that the total charge of the uniform background occupying the rhombus (shown in purple in Figure~\ref{fig:trian-lattice}) is opposite and equal in magnitude to the elementary charge of a particle. For the last lattice plane, half of such a rhombus is outside the lattice. The half of the height represents the squeezing of the droplet for an ideal triangular lattice and is given by the parameter $\delta$ as:
\begin{equation}
    \delta = \frac{h}{2} \quad\Rightarrow\quad \delta\sqrt{\bar{\rho}} = \frac{3^{1/4}}{2^{3/2}} \approx 0.465
\end{equation}
which is consistent with the large $\Gamma$ value of $\delta = 0.46\pm 0.01$ found from the simulations (see Figure~\ref{fig:squeezing}).

\subsection{Damping}
\label{sec:damping}

2D crystals are known not to have a truly long-range order. Rather, correlations of the translational order parameter decay slowly with distance, as a power law. Our crystal edge, however, does not present such algebraic order. Instead, the crystal signature of oscillations in the radial density profile decreases exponentially with the distance to the boundary. This should be understood as following from that the translational order is short-ranged in the fluid regime $\Gamma<\Gamma_m\approx 140$. Although the boundary is a strong perturbation, which fixes the position of the last plane of particles by the requirement of charge screening, such ordering is exponentially melted into the bulk.

Let us make this argument more quantitative. Starting from the crystal at the edge, the density can then be calculated from the theory of elasticity as
\begin{eqnarray} 
 \label{eq:rhoavg}
    \rho(r) 
    &=& \left\langle \sum_i \delta(r-R_i-u(R_i))\right\rangle_u 
    = \sum_i \int \frac{d^2q}{(2\pi)^2}e^{iq(r-R_i)-\bar{S}(q,R_i)}\,,
 \nonumber \\
    & & e^{-\bar{S}(q,R_i)} \equiv \left\langle e^{-iqu(R_i)}\right\rangle_u \,.
\end{eqnarray}
Here the second line introduces $e^{-\bar{S}}$ known as the Debye-Waller factor, $R_i$ are the equilibrium positions of lattice sites, $u(r)$ is the displacement field in the crystal and $\langle ... \rangle_u$ denotes the expectation value with respect to the linear elasticity Hamiltonian,
\begin{eqnarray}
    \langle A[u] \rangle_u &=& \frac{1}{Z}\int \mathcal{D}u\, A[u] \exp \left\{-\frac{\Gamma}{2}H[u]\right\}
 \nonumber \\
    H[u] &=& \int_M d^2r \frac{1}{2}C_{ijkl}u_{ij}u_{kl}\,.
 \label{harmonic-el}
\end{eqnarray}
Here $u_{ij}=\frac{1}{2}(\partial_i u_j + \partial_j u_i)$ is the symmetric strain tensor
and $C_{ijkl}$ is the elastic tensor of the crystal in units in which the charge of the particles is $e=1$, which was calculated at zero temperature in \cite{tkachenko1966stability,tkachenko1969elasticity}. For a triangular lattice $C_{ijkl}$ is highly symmetric, with only two independent components, the Lam\'e constants $\mu$ and $\lambda$. At zero temperature, $\lambda$ is divergent, and the longitudinal phonon modes are totally suppressed \cite{alastuey1981classical}. At finite temperature, however, $\lambda$ gets renormalized to some finite value (see \cite{morf1979temperature} for the related case of $1/r$ potential). Here the domain $M=D^2(0,R)$ is a disk of radius $R$ in $\mathbb{R}^2$. In translationally invariant systems the Debye-Waller factor $e^{-\bar{S}}$ is a function of the momentum $q$ only, but because of the presence of a boundary it gains a dependence on the lattice site. We are especially interested in how $\bar{S}(q,r)$ varies as $r$ moves from the boundary $r=R$ to the bulk of the crystal at a given temperature.

For simplicity, we consider the problem in the half-plane geometry, $M=\{(x,y)\in\mathbb{R}^2|x>0\}$. The radial density profile now corresponds to the density averaged over $y$ direction
\begin{eqnarray}
    \rho(x) &=& \int \frac{dy}{L_y} \sum_i \int \frac{d^2q}{(2\pi)^2}e^{iq(r-R_i)}\left\langle e^{-iqu(R_i)}\right\rangle_u \\
    &=& \frac{1}{L_y}\sum_i \int \frac{dq_x}{2\pi}e^{iq_x(x-X_i)}\left\langle e^{-iq_xu_x(R_i)}\right\rangle_u\,,
 \label{eq:rhoedgex}
\end{eqnarray}
where $R_i=(X_i,Y_i)$.

We proceed, heuristically, by  remembering that the fluctuations of the particles close to the boundary are small $u_x(X_i=0)\approx 0$. We also assume that there is no significant dependence of the average on $y$-coordinate and replace $u_x(R_i)\to u_x(X_i,0)-u_x(0,0)$ in (\ref{eq:rhoedgex}).  We obtain\footnote{There are two approximations involved here: (i) we substitute the soft wall boundary conditions of the 2D OCP by Dirichet boundary conditions on the elastic displacement field and (ii) we then use the infinite-system result for the correlation function (\ref{eq:DW100}). Corrections are not expected to change the leading exponential decay of (\ref{eq:DW100}), and we discuss improvements to these approximations in section \ref{sec:conclusions}.}
\begin{eqnarray}
    \rho(x) &=&  \sum_{X_i} \int \frac{dq_x}{2\pi}e^{iq_x(x-X_i)}\left\langle e^{-iq_x(u_x(X_i,0)-u_x(0,0))}\right\rangle_u \,.
 \label{eq:rhoedgex2}
\end{eqnarray}
The summation in (\ref{eq:rhoedgex2}) is performed over crystal planes $X_i$. 

The Debye-Waller factor is now given by 
\begin{equation}
    e^{-\bar{S}(q_x,X)}=\left\langle e^{-iq_x(u_x(X,0)-u_x(0,0))}\right\rangle_u
 \label{eq:DW100}
\end{equation}
with the average computed in linear elasticity theory. The correlation function (\ref{eq:DW100}) decays as a power law in the 2D crystal phase but is expected to decay exponentially above melting point. For large enough $X$ the rate of this exponential decay is controlled by the bulk correlation function and up to pre-exponential factors
\begin{equation}
    \langle e^{-iq_x(u_x(X,0)-u_x(0,0))}\rangle\approx e^{-\frac{q_x^2}{2}\langle (u_x(X,0)-u_x(0,0))^2\rangle}
    \sim e^{-\frac{q_x^2}{G^2}\frac{X}{\xi}}
    \,.
    \label{eq:harmonicapp2}
\end{equation}
It is convenient to normalize the exponent by the principal reciprocal wave vector $G=\frac{2\pi}{h}$ so that the correlation length $\xi$ has the units of length. 

Substituting the approximation (\ref{eq:harmonicapp2}) back in (\ref{eq:rhoedgex2}) then gives
\begin{eqnarray}
    \rho(x) &=& \sum_{X_i} \int \frac{dq_x}{2\pi}e^{iq_x(x-X_i)} e^{-\frac{q_x^2}{G^2}\frac{X_i}{\xi}}
    = \sum_{X_i}\frac{1}{\sqrt{2\pi\sigma^2}}e^{-\frac{1}{2}\left[\frac{(x-X_i)}{\sigma(X_i)}\right]^2},
    \label{eq:rhox}
\end{eqnarray}
which is quite an intuitive result: the delta peaks corresponding to the lattice planes get smeared by thermal fluctuations into Gaussians. The width of these Gaussians increases with the distance from the boundary,
\begin{equation}
    G^2\sigma^2(X)=\frac{2X}{\xi},
    \label{eq:sigmalinear}
\end{equation}
which means that, at large $X$, $\sigma$ is much larger than the inter-particle distance and one recovers the constant density fluid bulk profile. Moreover, the argument goes on to explain why the damping of the observed oscillations is exponential. Let us perform a summation over lattice planes $X_i=nh$ on the expression for $\rho(x)$, assuming that $x$ is sufficiently deep into the bulk of the fluid
\begin{eqnarray}
    \rho(x) &=& \sum_{n=0}^{\infty}\frac{1}{\sqrt{2\pi\sigma^2}}e^{-\frac{(x-h n)^2}{2\sigma^2}} \approx \sum_{n=-\infty}^{\infty}\frac{1}{\sqrt{2\pi\sigma^2}}e^{-\frac{(x-h n)^2}{2\sigma^2}}
 \label{eq:rhoexp3} \\
    &=&\frac{1}{h}\sum_{m=-\infty}^{\infty}e^{-\frac{1}{2}\left(\frac{2\pi\sigma}{h}\right)^2 m^2+i\frac{2\pi m}{h}x} 
 \nonumber \\
 	&=& \frac{1}{h}\left(1+2e^{-2\pi^2\sigma^2/h^2}\cos(Gx)+\dots\right)\,.
 \label{eq:rhopoisson}
\end{eqnarray}
Here we expanded the summation to the infinite region in the first line using $x\gg \sigma$. In the second line the Poisson's summation formula was used. 
Sufficiently deep in the bulk ($x$-large) the combination $2\pi^2\sigma^2/h^2=4\pi^2 x/(\xi G^2 h^2)=x/\xi$ becomes larger than one and the second term in the expansion (\ref{eq:rhopoisson}) dominates the oscillating corrections to the background density. This term gives oscillations of period $h$, modulated by an envelope $e^{-2\pi^2\sigma^2/h^2}=e^{-x/\xi}$ over the constant bulk value, which explains the observed exponential decay of the oscillations. The correlation length for the decay of density oscillations coincides with the bulk correlation length defined by displacements as  (\ref{eq:harmonicapp2}).

\subsection{Comparison with numeric results}
\label{sec:comparison}

\begin{figure}[H]
  \centering
    \includegraphics[width=\textwidth]{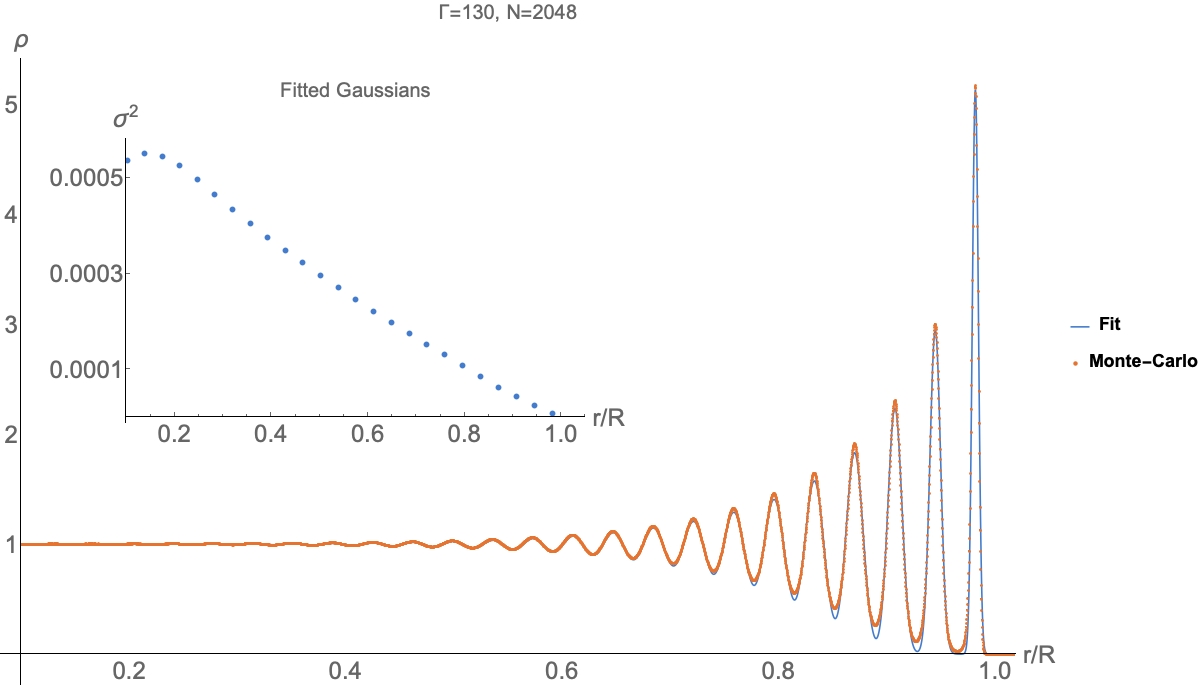}
  \caption{Density profile of the 2D OCP at $\Gamma=130$, $N=2048$. The points are a Monte Carlo simulation while the solid line is given by (\ref{eq:rhoexp3}) where the width of the gaussian peaks varies with the distance to the boundary according to (\ref{eq:sigmalinear}). Aside from the first two peaks, we see very good agreement.  }
  \label{fig:sigmalinear}
\end{figure}

To compare with numerics, we fit the density profile with the full expression (\ref{eq:rhoexp3}). Figure \ref{fig:sigmalinear} shows a comparison for $\Gamma=130$, $N=2048$. Aside from the first two peaks, we see very good agreement, and the inset shows how the thermal fluctuations of the lattice planes do increase linearly with the distance from the boundary as predicted by equation (\ref{eq:sigmalinear}).

\subsection{Divergence of the Correlation Length}
\label{sec:correlation}

In figure \ref{fig:peaks} we showed that the damping of the density oscillations becomes smaller at large $\Gamma$. After seeing the relationship between the damping and the correlation functions of the translational order parameter of the crystal, an explanation for the temperature dependence of the damping follows, at least close to $\Gamma_m\approx 140$. One of the signatures of the crystallization transition is the divergence of the correlation length of the order parameter. The above discussion shows that the damping of oscillations in the radial density profile is a probe into this effect, and should show the same divergence.

\begin{figure}[H]
  \centering
    \includegraphics[width=0.7\textwidth]{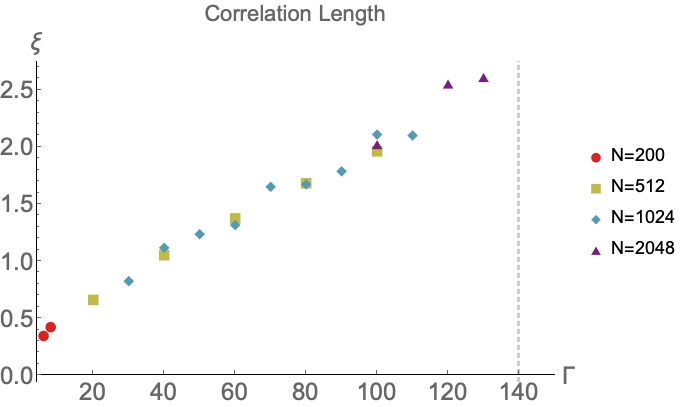}
  \caption{The correlation length extracted from the decay of the oscillations in the radial density profile, for different values of $\Gamma$ and $N$. We do not see indications of divergence of $\xi$ up to $\Gamma=130$. Larger system sizes are necessary to study the temperatures closer to the phase transition at $\Gamma=\Gamma_m\approx 140$ (vertical line) and to make any conclusions on the nature of the melting transition.}
  \label{fig:xinew}
\end{figure}

In figure \ref{fig:xinew} we plot the correlation length $\xi$ extracted from density profiles as a function of $\Gamma$. As expected the correlation length monotonously increases with $\Gamma$. The phase transition to the crystal phase is expected at $\Gamma=\Gamma_m\approx 140$. There are no strong signs of divergence of $\xi$ up to $\Gamma=130$. For $\Gamma$ closer to the transition value, the oscillations persist further into the bulk, so that it is necessary to have a larger system to ensure that the whole bulk/boundary interface is visible and finite-size effects are suppressed. Unfortunately, already at $\Gamma=130$, there are indications that the number of particles $N=2048$ is not sufficient and one should go to even larger system sizes to be able to analyze the behavior of the correlation length close to the phase transition. This analysis would allow one to test whether a second-order phase transition is realized or a weak first-order transition happens instead. We conclude this section with a remark that better control of finite size effects can be achieved in studying 2d OCP in cylindrical geometry, the point elaborated in the next section.

\section{Discussion and Conclusions}
 \label{sec:conclusions}

In this work we performed extensive Monte-Carlo simulations of the 2D OCP. There is a wide range of temperatures $2<\Gamma<140$ at which the density of the 2D OCP exhibits oscillations at wave vectors of the order of an inter-particle distance, besides other boundary features. These numerical observations (see Sec.~\ref{sec:MC}) support the ``freezing at the edge'' picture. In this picture the bulk plasma is in a liquid state while the crystal order is formed at the edge of the droplet. In particular, we showed that the wave vector of oscillations of the boundary density, which are very pronounced at $2\ll \Gamma$, is in a very good agreement with the lattice constant of the triangular lattice even at $\Gamma<\Gamma_m\approx 140$, i.e., in the regime which corresponds to the liquid in the bulk.

Let us remark here that while historically the disk has been a favorite droplet geometry in 2D OCP studies, other geometries can be considered and sometimes give more control. For example, one can consider the 2D OCP on the surface of a cylinder rolled in $y$ direction $y\in [0,2\pi r_0]$ (see \ref{app:cylinder}). In this geometry the droplet has a shape of a section of the cylinder with the longitudinal coordinate $x$ roughly in the range $x\in [-L,L]$. This ``droplet'' has two boundaries. The advantage of this geometry is that the geodesic curvature of the boundaries is zero and, more importantly, does not change with the increased number of particles as in the case of the disk. Therefore, one can fix $r_0$ and increase the number of particles $N$, increasing the available range for the distance to the boundary $x$ linearly with $N$ (vs. $\sqrt{N}$ for the disk). Moreover, for the tuned values of $r_0$ one can wrap the triangular lattice around the cylinder without introducing any lattice defects, which is not possible for the disk.

\begin{figure}[H]
  \centering
    \includegraphics[width=0.6\textwidth]{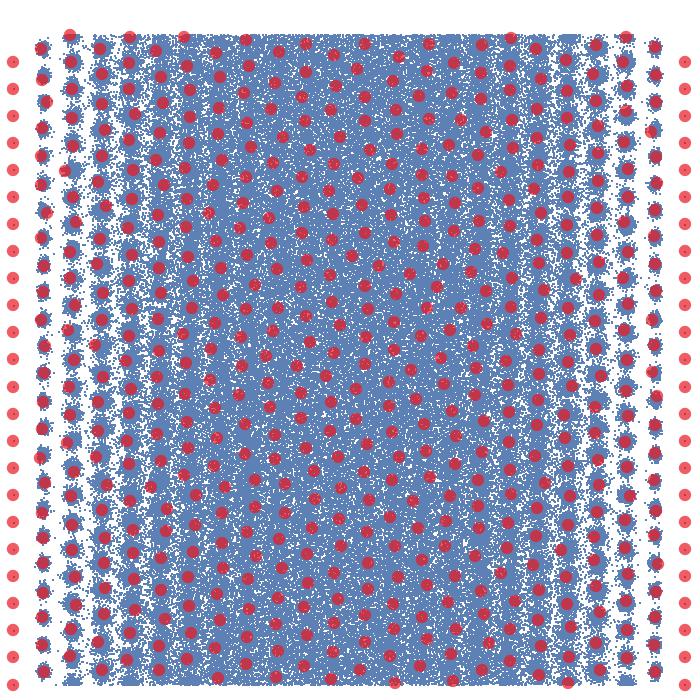}
  \caption{Blue dots are the positions of 576 particles for 500 Monte-Carlo configurations on top of each other. The temperature is $\Gamma=120$. The simulation is done on a cylinder with the vertical axes showing the angle (compact direction). The leftmost and the rightmost columns of red dots correspond to fixed particle positions. The remaining red dots show the lowest energy configuration encountered during the Monte-Carlo simulation. }
  \label{fig:beta50-snapshot}
\end{figure}

In figure \ref{fig:beta50-snapshot} we present the results of a Monte-Carlo run we performed in the geometry of a cylinder  (for definitions of the OCP on the cylinder see \ref{app:cylinder}). We placed $N=576=24^2$ particles on the surface of the cylinder fixing the positions of the 24 leftmost and 24 rightmost particles (shown as a leftmost and rightmost columns of red dots, respectively) so that it might be possible in principle to put all other particles in between them as a perfect triangular lattice array. Then we show 500 typical configurations at $\Gamma=120$ allowing all particles to move except for the fixed ones. All positions of particles for all snapshots are put on top of each other as blue dots in figure \ref{fig:beta50-snapshot}. The best energy configuration of particles encountered during all the runs is shown as red dots. One can notice that it is not a perfect triangular lattice but is pretty close, showing that we had sufficiently many runs to explore configurations near the absolute energy minimum. We would like to attract the reader's attention to the following features clearly seen in this figure. (i) the blue dots covering is almost uniform deep in the bulk, consistent with the liquid bulk state at $\Gamma=120$, (ii) a good crystal can be seen near the left and right boundaries of the droplet, (iii) the increased smearing of the crystal towards the bulk (iv) anisotropy in the fluctuations of the positions of particles close to the boundaries, seen as elliptic shapes formed by the blue dots in this region, (v) the anisotropy seems to get smaller deeper in the bulk.

The features of figure \ref{fig:beta50-snapshot} strongly support the ``freezing at the edge'' picture for 2D OCP for $\Gamma \gtrsim 10$ advocated in this work. The density profile (\ref{eq:rhox}) with (\ref{eq:sigmalinear}) is in good agreement with the Monte Carlo data. To have a more quantitative analysis of the data one needs to replace the partially heuristic arguments used in section \ref{sec:damping} with a more precise derivation of the boundary density profile due to the boundary perturbation in the 2d melting problem. This can be done more rigorously in the setting in figure \ref{fig:beta50-snapshot}, which involves Dirichlet boundary conditions rather than the soft wall boundary conditions analysed in this paper. We leave this and many other interesting questions for future work.

\section{Acknowledgments}

We are grateful to Paul Wiegmann, Leo Radzihovsky, and Robert Konik for useful discussions and for multiple comments regarding this manuscript. 
This research was supported by grants NSF DMR-1606591 (AGA, GC) and US DOE DESC-0017662 (AGA). 


\clearpage

\appendix

\section[\mbox{}\hspace{2cm} \texorpdfstring{$1/N$}{} expansion for the boundary density at \texorpdfstring{$\Gamma=2$}{}]{$1/N$ expansion for the boundary density at \texorpdfstring{$\Gamma=2$}{}}
 \label{app:G2boundary}

The density for $N$ particles at $\Gamma=2$ is given by (see, for example, section 2.1 in \cite{jancovici1982classical})
\begin{eqnarray}
    \rho(r) = e^{-\pi r^2} \sum_{n=0}^{N-1} \frac{(\pi r^2)^n}{n!} = \frac{\Gamma(N,\pi r^2)}{\Gamma(N)}\,.
\end{eqnarray}
Here we use units in which the background density is $\bar\rho=1$. We introduce $\xi=\frac{r^2-R^2}{2R^2}$ and rewrite
\begin{eqnarray}
    \hat\rho(\xi)\equiv \rho(r)-\rho_0(r) &=& e^{-\pi R^2(2\xi+1)} \sum_{n=0}^{N-1} \frac{(\pi R^2(2\xi+1))^n}{n!} -\theta(-\xi)\,.
\end{eqnarray}
We extend the $\hat\rho(\xi)$ so that for $\hat\rho=0$ for $\xi\leq-1/2$ and take a Fourier transform
\begin{eqnarray}
    \hat\rho_k &=& \int_{-\infty}^{+\infty}d\xi\, \hat\rho(\xi) e^{-ik\xi}
 \\
    &=& e^{i\frac{k}{2}}\left(\frac{1}{2\pi R^2}\sum_{n=0}^{N-1}\left(1+i\frac{k}{2\pi R^2}\right)^{-n-1}+\frac{e^{-i\frac{k}{2}}-1}{ik}\right) 
 \\
    &=& \frac{1}{ik}\left[1-e^{i\frac{k}{2}}\left(1+i\frac{k}{2N}\right)^{-N}\right]
 \\
    &=& \frac{1}{ik}\left[1-e^{i\frac{k}{2}-N\ln\left(1+i\frac{k}{2N}\right)}\right]\,.
\end{eqnarray}
Expanding in $1/N$ we have
\begin{eqnarray}
    \hat\rho_k &=& \frac{1}{ik}\left[1-e^{-\frac{k^2}{8N}+i\frac{k^3}{24 N^2}+\frac{k^4}{64 N^3}+\ldots}\right]
 \nonumber \\
    &=& -i\frac{k}{8N}+\frac{k^2}{24N^2}+i\frac{k^3}{128N^2}\left(1+\frac{2}{N}\right)+\ldots\,.
 \label{eq:sing1N}
\end{eqnarray}
Taking the inverse Fourier transform of this expression then gives equation (\ref{exp10}),
\begin{eqnarray}
    \hat\rho(\xi) = -\frac{1}{8N}\delta'(\xi)-\frac{1}{24N^2}\delta''(\xi) -\frac{1}{128N^2}\left(1+\frac{2}{N}\right)\delta'''(\xi)+\ldots \,.
 \label{eqdelta}
\end{eqnarray}

We notice that alternatively we could have used a different $1/N$ expansion,
\begin{eqnarray}
    \hat\rho_k &=& \frac{1}{ik}\left[1-e^{-\frac{k^2}{8N}+i\frac{k^3}{24 N^2}+\frac{k^4}{64 N^3}+\ldots}\right]
 \nonumber \\
    &=&  \frac{1}{ik}\left[1-e^{-\frac{k^2}{8N}}\right] -e^{-\frac{k^2}{8N}}\left(\frac{k^2}{24 N^2}-i\frac{k^3}{64 N^3}+\ldots\right)\,,
\end{eqnarray}
keeping $k^2/N$ in the exponent. In this case the typical values of $k$ are of the order of $\sqrt{N}$. This is the scale of the boundary smearing, which is lost in the singular expansion (\ref{eq:sing1N},\ref{eqdelta}).

\section[\mbox{}\hspace{2cm}  Sum rules]{Sum rules}
 \label{app:sumrules}

Sum rules have a long history in statistical mechanics. For a review of sum rules for charged fluids, and OCP in particular, see Refs.~\cite{martin1988sum},\cite{kalinay2000sixth}. More recently a combination of exact sum rules (or more generally Ward identities) with $1/N$ expansion was shown to be a powerful tool in studying the OCP \cite{zabrodin2006large}. Here, to make this work more self-contained we present a derivation of the simplest sum rule, which was used in section \ref{sec:sumrules}. 

It is straightforward to derive the following useful sum rules for the system (\ref{ZOCP}) 
\begin{eqnarray}
    \int d^2x\,\rho &=&N\,,
 \label{sr0} \\
    \frac{1}{N}\int d^2x\, \rho r^2  &=& \frac{R^2}{2}-\frac{\Gamma-4}{2\pi \Gamma}\,.
 \label{sr1}
\end{eqnarray}
The first one is just a statement that the total number of particles is fixed and equal $N$. The second one is less trivial and can be derived in the following way. Let us make a change of variables in the partition function (\ref{ZOCP}) $z_j\to \lambda z_j$. As this is just a change of integration variables the partition function (\ref{ZOCP}) will not change, which is equivalent to the following statement
\begin{eqnarray}
    \lambda^{2N} \lambda^{\Gamma \frac{N(N-1)}{2}} \left\langle \exp\left(-\Gamma\sum_{j=1}^N \Big[W(\lambda z_j,\lambda\bar{z}_j) -W(z_j,\bar{z}_j)\Big]\right) \right\rangle = 1\,.
 \label{sr10}
\end{eqnarray}
Here the first factor comes from the measure of integration $\prod_jd^2z_j$, the second from the logarithmic interaction and the last factor is from the change of the background potential. The average is performed with (\ref{ZOCP}). Assuming that $\lambda=1+\epsilon$, expanding (\ref{sr10}) in $\epsilon$, and keeping only the terms of the first order in $\epsilon$ we obtain:
\begin{eqnarray}
    2N +\Gamma \frac{N(N-1)}{2} -\Gamma\left\langle\sum_{j=1}^N\left(z_j\frac{\partial W}{\partial z_j}+\bar{z}_j\frac{\partial W}{\partial \bar{z}_j}\right)\right\rangle = 0\,.
 \label{sr20}
\end{eqnarray}
For the potential (\ref{Wharm}) we rewrite (\ref{sr20}) as 
\begin{eqnarray}
     \frac{1}{N}\left\langle\sum_{j=1}^N |z_j|^2\right\rangle = \frac{N}{2\pi} -\frac{\Gamma-4}{2\pi\Gamma}\,.
 \label{sr30}
\end{eqnarray}
The exact sum rule (\ref{sr30}) can be rewritten as (\ref{sr1}) with the density defined by (\ref{eq:density}) and $N=\pi R^2$.

It is convenient to rewrite (\ref{sr1}) as the moment of the density distribution relative to $\rho_0(r)$ from (\ref{rho0r})
\begin{eqnarray}
    \frac{1}{N}\int d^2x\, \Big(\rho-\rho_0\Big) r^2  &=& -\frac{\Gamma-4}{2\pi \Gamma}\,.
 \label{sr40}
\end{eqnarray}

Let us define the following quantity
\begin{eqnarray}
    d &\equiv \frac{1}{4N}\int d^2x\, (\rho-\rho_0) r^2  
    = -\frac{1}{8\pi}\left(1-\frac{4}{\Gamma}\right)\,.
 \label{dsumr}
\end{eqnarray}
Here the second equality follows from the sum rule (\ref{sr40}). Following \cite{can2014edgelaughlin} we can connect $d$ defined by (\ref{dsumr}) to the dipole density at the boundary in large $N$ limit. Indeed, we have
\begin{eqnarray}
    d &\equiv \frac{1}{4N}\int d^2x\, (\rho-\rho_0) r^2
    =\frac{1}{4N}\int d^2x\, \hat\rho (r^2-R^2)\,,
 \label{ddef20}
\end{eqnarray}
where we used the fact that the integral of $\hat\rho\equiv\rho-\rho_0$ is zero by (\ref{sr0}). Assuming that $\hat\rho(r)$ is non-vanishing only close to the boundary, we proceed as
\begin{eqnarray}
    d \approx \frac{1}{4N}\int dr\,2\pi R \, \hat\rho\, (r-R)2R = \int dr\, \, \hat\rho\, (r-R)\,.
\end{eqnarray}
The last expression defines the boundary density of the dipole moment relative to $\rho_0(r)$. We remark here that the sum rule (\ref{dsumr}) is exact, while the connection of the quantity $d$ to the boundary dipole density becomes exact only in the large $N$ limit.

For numerical checks of the sum rule it is convenient to rewrite the expression (\ref{dsumr}) as 
\begin{eqnarray}
    \frac{\Gamma}{4}\Big(8\pi d+1\Big)=1\,.
\end{eqnarray}
Here the left hand side can be computed from Monte Carlo data and the microscopic expression for $d$:
\begin{eqnarray}
     d=\frac{1}{4N}\left\langle\sum_{j=1}^N |z_j|^2\right\rangle - \frac{N}{8\pi}\,,
 \label{dmicro}
\end{eqnarray}
which is equivalent to (\ref{ddef}).

\section[\mbox{}\hspace{2cm}  2D OCP in cylinder geometry]{2D OCP in cylinder geometry}
 \label{app:cylinder}


There is a simple generalization of the problem studied in this paper, that of the one-component plasma in a surface. Again we consider $N$ particles interacting through a repulsive Coulomb potential $v(\mathbf{x},\mathbf{x'})$ in the presence of a potential $W(\mathbf{x})$ generated by a background of constant opposite charge. The total energy is
\begin{eqnarray}
    E &=& q^2\sum_{1\leq j<k\leq N}v(\mathbf{x}_i,\mathbf{x}_j) + q^2\sum_{j=1}^N W(\mathbf{x}_j)\,,
\end{eqnarray}
and the partition function is
\begin{eqnarray}
    Z &=& \int \prod_{j=1}^N d^2x_j \; \exp\left\{-\Gamma \left[\sum_{1\leq j<k\leq N}v(\mathbf{x}_i,\mathbf{x}_j) +  \sum_{j=1}^N  W(\mathbf{x}_j)\right]\right\}\,.
 \label{Zcyl}
\end{eqnarray}
Here the Coulomb potential $v(\mathbf{x},\mathbf{x'})$ is defined by the Green's function of the Laplace-Beltrami operator\footnote{For details on regularization of the Green's function on general surfaces see Ref.~\cite{bogatskiy2019vortex}.} of the surface,
\begin{equation}
    \Delta v(\mathbf{x},\mathbf{x'}) = -2\pi\delta(\mathbf{x}-\mathbf{x'}).
\end{equation}
Likewise, the background potential $W(\mathbf{x})$ satisfies
\begin{equation}
    \Delta W(\mathbf{x}) = 2\pi.
\end{equation}

In the plane, one recovers the logarithmic Coulomb potential and the quadratic background potential, thus recovering (\ref{EOCP}) and (\ref{ZOCP}). In the cylinder geometry, one has similar expressions. Let $\eta$ be the coordinate along the cylinder axis and $\phi$ the azimuth angle, around the cylinder. In these coordinates,
\begin{equation}
    v(\mathbf{x},\mathbf{x'})=-\ln\left[2\cosh\left(\frac{\eta-\eta'}{r_0}\right)-2\cos(\phi-\phi')\right]^{1/2},
 \label{intcyl}
\end{equation}{}
where $r_0$ is the radius of the cylinder. This reduces, at distances much smaller than the radius of the cylinder $r_0$, to the planar potential $v=-\ln |\mathbf{x}-\mathbf{x'}|/r_0$ and, at distances much larger than $r_0$, to the one-dimensional Coulomb potential $v=-|\eta-\eta'|/2r_0$. The background potential is, in these coordinates,
\begin{equation}
    W(\mathbf{x})=W(\eta,\phi)=\pi \eta^2\,.
 \label{Wcyl}
\end{equation}
The results in figure \ref{fig:beta50-snapshot} are presented for the model (\ref{Zcyl}) with (\ref{intcyl},\ref{Wcyl}) and fixed particle positions at the boundaries of the cylindrical droplet.

\section*{References}

\bibliography{coulomb-droplet}

\bibliographystyle{unsrt}

\end{document}